\documentclass[twocolumn,superscriptaddress,amsmath,amssymb,prb]{revtex4}

\usepackage{graphicx,tabularx}
\usepackage[english]{babel}

\begin{document}

\title{Shaping single walled nanotubes}

\author{A. Zobelli}
\affiliation{Laboratoire de Physique des Solides, Univ. Paris-Sud, CNRS, UMR 8502, F-91405 Orsay Cedex, France}

\affiliation{Technische Universit\"{a}t Dresden,
Instit\"{u}t f\"{u}r Physikalische Chemie und Elektrochemie, D-1062 Dresden,
Germany}

\author{A. Gloter}
\affiliation{Laboratoire de Physique des Solides, Univ. Paris-Sud, CNRS, UMR 8502, F-91405 Orsay Cedex, France}

\author{C. P. Ewels}
\affiliation{Universit{\'e} de Nantes, 
Nantes Atlantique Universit{\'e}s, CNRS,
Institut des Mat{\'e}riaux Jean
Rouxel (IMN), UMR6502, BP32229, F-44322 Nantes Cedex 3, France}

\author{C. Colliex}
\affiliation{Laboratoire de Physique des Solides, Univ. Paris-Sud, CNRS, UMR 8502, F-91405 Orsay Cedex, France}

\begin{abstract}

We show that electron irradiation in a dedicated scanning transmission microscope can be used as a nano-electron-lithography technique allowing the controlled reshaping of single walled carbon and boron nitride nanotubes.
The required irradiation conditions have been optimised on the basis of total knock-on cross sections calculated within density functional based methods. 
It is then possible to induce morphological modifications, such as a change of the tube chirality, by removing severals tens of atoms with a nanometrical spatial resolution.
These irradiation techniques could open new opportunities for nano-engeneering a large variety of nanostructured materials.

\end{abstract}

\maketitle

\section{Introduction}

Irradiation of nanotubes is now a widespread issue which is primarily encountered within two contexts. 

Firstly, nanotube irradiation with different energetic particles, such as $\gamma$ rays\cite{Hulman-05,Skakalova-04}, electrons\cite{Beuneu-99,Kis-04}, protons\cite{Basiuk-02,Khare-03} and ions\cite{Gomez-05} can be deliberately used to alter the chemical, mechanical and electronic properties of the tubes. For example  Kis and coworkers\cite{Kis-04} have shown a strong stiffening of bundles of carbon nanotubes after electron irradiation. Another example can be found in the work of Gomez-Navarro and coworkers\cite{Gomez-05} where Ar$^+$ ion bombardment of carbon nanotubes provoked a dramatic increase in the tube electrical resistivity. 

Secondly, irradiation is also an unavoidable side effect occurring when highly energetic particles are used to investigate structural and spectroscopic properties of the tubes. This is of particular importance for transmission electron microscopy (TEM) which 
allows the observation of individual defects on nanotubes\cite{Zobelli-06,Hashimoto-04,Suenaga-07}.  Meanwhile the electron beam might also damage the nanotube structure.  This eventually leads to extended wavy morphologies, and ultimately complete tube amorphisation\cite{Ajayan-93}. Recently, Yuzvinsky and coworkers\cite{Yuzvinsky-06} demonstrated that the crystallinity of the tube can be preserved through thermal treatments during the TEM observation. 

A more limited number of studies have tried to use the focussing properties of electrons to locally modify the nanotube structure. Li and Banhart\cite{Li-04} have demonstrated that focusing electron probes on multiwalled carbon nanotubes (MWNT) can bend the tubes or locally produce carbon onions from the tubes walls. Yuzwinsky and coworkers\cite{Yuzwinsky-05} have demonstrated that MWNT tubes can be cut with a SEM even at low electron voltage when a degraded vacuum is present. 

Nonetheless, the technological potential of electron irradiation of nanotubes is far from being realized. This is primarily because, until now, theory was not able to predict quantitatively the expected defect structures as a function of the irradiation parameters, and experiments were not performed with sufficient spatial control.

In the current paper we present a fundamental improvement in the irradiation techniques, optimising irradiation conditions on the basis of calculated knock-on cross sections.\cite{Zobelli-07-2} We show that dislocations of a few nanometers in length, corresponding to the removal of a few tens of atoms, can be obtained with nanometrical accuracy using subnanometrical focused probes in a dedicated scanning transmission electron microscope (STEM). Experimental shaping of single walled carbon and boron nitride nanotubes are then obtained demonstrating that chiralities of the tube can be locally changed with nanometric spatial control. 

\section{Results}

Under electron irradiation, defects appear in single walled nanotubes mainly due to direct (quasi-elastic scattering) collisions between the relativistic electrons of the beam and the atomic nucleus. Removal of the atom is thus obtained through the so-called knock-on effect. 
In Ref.\cite{Zobelli-07-2} the theory of irradiation processes is detailed and total knock-on cross sections are presented for perfect carbon and boron nitride nanotubes as a function of the energy of the incident beam for different emission sites around the tube circumference. We note however that these calculations were only performed for the creation probability of single isolated vacancies. 
Under electron irradiation it has been reported that nanotubes show kinks or extended defect formation and these structures are interpreted as the creation of dislocations by sequential removal of a series of adjacent atoms. The preferential creation of dislocation lines versus a random distribution of single vacancies has been explained by a laddering mechanism: once a primary vacancy is formed, the emission probability for one of the atoms neighboring the vacancy is higher than for atoms in a perfect graphitic environment.\cite{Zobelli-06,Niwase-05,Ding-07-1,Ding-07-2,Kotakoski-06} This process has been proposed on the basis of lower vacancy formation energies at atom sites neighboring a pre-existing vacancy, or located at the ends of a dislocation line. However, vacancy formation energies are only indirectly related to emission probabilities of atoms. Consequently, we have derived the total knock on cross section for atoms emitted during the first steps of the dislocation propagation process. As previously reported\cite{Zobelli-07-2} our knock-on cross section calculation involves the derivation of the emission energy threshold through extended density functional based  molecular dynamics simulations\cite{Krasheninnikov-05,Zobelli-07-2} and a successive integration of the Mott cross section over the allowed emission solid angle.

\begin{figure}[tbhp]
\centering
\includegraphics[width=\columnwidth]{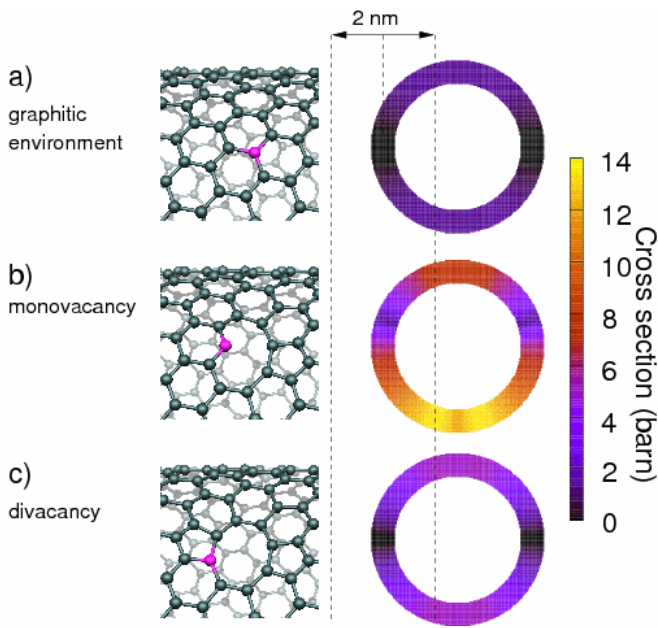}
\caption{Right part of the figure: total knock-on cross section as a function of the position on a tube section  of an atom in a perfect graphitic environment (a),  of a double coordinated atom neighboring a monovacancy (b) of an atom neighboring a divacancy (c). The beam incidence direction is aligned along the figure's vertical axis. The two vertical dashed lines represent the irradiated zone used to obtain experimental tube shaping.
Left section: structures of the targeted nanotubes. Emitted atoms are marked in magenta.}
\label{crosss}
\end{figure}

Figure \ref{crosss} shows the knock-on process for, respectively, an atom in a perfect graphitic environment (a), a doubly coordinated atom neighboring a monovacancy (b) and an atom neighbouring a divacancy (c). The right side graphs represent the transversal section of the nanotube where the color scale refers to the total knock on cross section at that location for each of these atom types.
While calculations were performed for a wide range of incoming electron energies (see Sec. I in supporting materials), Figure \ref{crosss} only presents the results obtained for an electron beam energy 20 keV above the threshold energy at which defects can be generated. Indeed, experimental tube shaping has been done at this corresponding energy.

For a perfect carbon nanotube (Figure \ref{crosss}.a) a strong angular dependence on the emission probability is obtained. The cross section decreases with increasing angle between the beam incidence direction and the normal to the tube, and a forbidden emission region appears corresponding to the side walls of the tube.
Once a first vacancy is generated a pentagonal ring appears with the reconstruction of a C-C bond between two vacancy neighbors.  The third neighboring atom remains with one dangling bond and shifts slightly radially outwards. For this lower coordinated atom knock-on cross sections are almost one order of magnitude higher than for an atom in a perfect tube (Figure \ref{crosss}.b): on tube regions perpendicular to the electron beam a maximum cross section of 13.4 barn is obtained while a perfect graphitic environment shows only cross sections of about 1.4 barn.
The asymmetry between the upper and lower part of the tube in figure \ref{crosss}.b is due to the outward movement of the doubly coordinated atom, which makes emission into the tube cavity more difficult.

After removal of one atom close to a preexisting vacancy, the nanotube relaxes with the formation of a large divacancy and the creation of two pentagonal rings. The cross section map reported in figure \ref{crosss}.c corresponds to the emission of one of the atoms neighboring a divacancy. A maximal cross section of 5.2 barn is found in the tube sections normal to the beam incident direction, showing a partial stabilization of the atom compared to the previous case. 
With further knock-on events, odd and even numbers of vacancies are sequentially obtained and form a dislocation line.  For odd numbers of vacancies the tube relaxes similarly to the monovacancy case, with the appearance of a pentagonal ring and a doubly coordinated atom. This is equivalent to a basal plane dislocation terminating in the shuffle plane. Analogously the removal of an even number of atoms gives a structure with ends topologically equivalent to a divacancy, i.e. two pentagonal rings at the ends of the dislocation line. In this case both dislocation cores terminate in the glide plane. Due to these morphological similarities we expect the emission knock-on cross section for the odd and even cases to correspond to values close to those obtained for the monovacancy and divacancy respectively. The high emission probabilities for atoms at the two ends of a dislocation compared to atoms in a perfect graphitic environment support the existence of preferential sites for atomic emission that provoke the propagation of dislocation lines under electron irradiation through a laddering mechanism. One can see direct experimental confirmation of this theoretical model in the TEM movies presented in a recent work of Suenaga \textit{et al.} (Ref. \cite{Suenaga-07}, supporting materials movie S4). The movie shows an initial short dislocation which ends at two kinks on the tube side walls. During the acquisition time the two kinks move far from each other with a related tube diameter reduction. This behaviour corresponds to atom emission under irradiation at the two ends of the dislocation loop, which propogates the dislocation along the tube axis.

\begin{figure*}[bthp]
\centering
\includegraphics[width=\linewidth, bb=0 0 654 237]{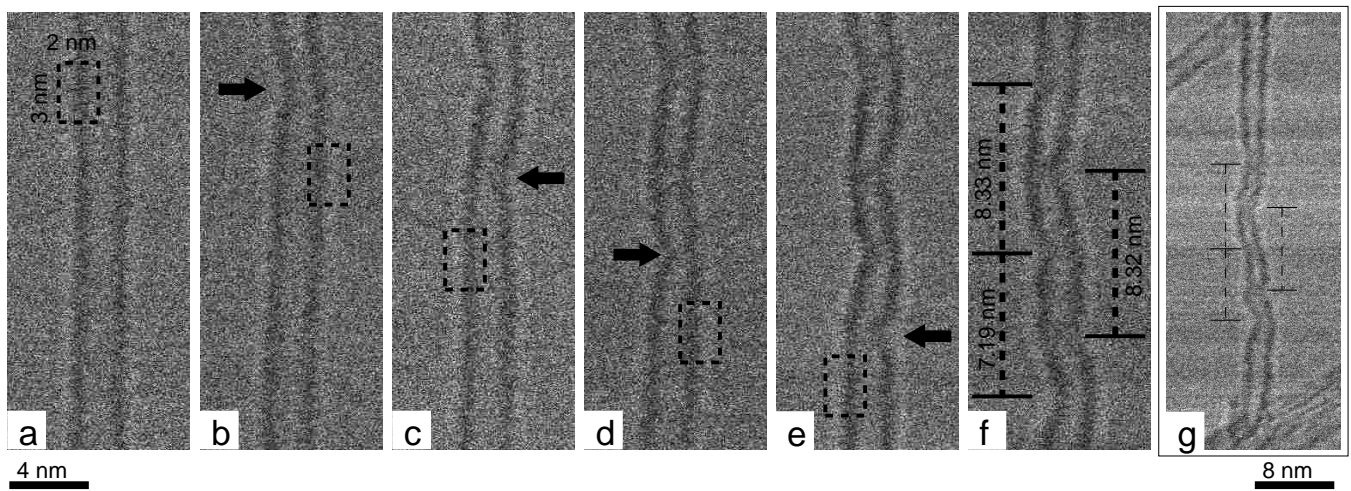}
\caption{Local electron irradiation of a single walled carbon nanotube. Irradiation zones of $2 \times 3$ nm$^2$ are represented by the dashed rectangles. (a) Original tube with a diameter of 2.4 nm and an apparent perfect crystallinity. (b) After the first irradiation cycle the tube shows a kink associated with a slight bending (black arrow) in direct correspondence with the chosen, irradiated zone. (c-f) Similar defective structures appear after each irradiation cycle as noted with arrows. (g) Final structure observed at a lower magnification demonstrating that the non irradiated zones at the two extreme parts of the tubes remain unaltered.}
\label{kinkseries}
\end{figure*}

Experimental conditions for single walled carbon nanotube irradiation have been optimized on the basis of the calculated total knock-on cross sections in order to obtain tube shaping capability. In figure \ref{kinkseries} such a shaping of a single walled carbon nanotube is obtained by successive cycles of local electron irradiation. Five extended kinks have been obtained sequentially, on alternating sides of the tube. 
To obtain such controlled irradiation, we select illuminated area, electron beam current density and electron beam energy in order to have a low atom ejection rate during the exposure time. Irradiation cycles were performed using an electron beam energy of 100 keV
(roughly 20 keV above threshold voltage\cite{Smith-01}),
an electron current of $\sim 140$ pA and exposure times of 60 s. The beam convergence half-angle has been set to about 7.5 mrad which corresponds to an electron probe diameter of 0.8 nm. Five irradiation cycles were performed, choosing sequential irradiation zones on alternating sides of the tube at about 4 nm spacing along the tube axis. The scanning regions, represented in figure \ref{kinkseries} by the dashed rectangles, are limited to a $ 2 \times 3$ nm$^2$ area and partially illuminate the external section of the tube walls. As shown in figure \ref{crosss}, atoms located on these sectors of the tube have the lowest total knock-on cross section and therefore a better control of the irradiation process can be obtained using longer exposure times. On the basis of the experimental irradiation conditions and the calculated cross section, a low number of primary single vacancies ($\sim 3$) can be generated during the exposure time (details of the calculation are given in Sec. II of the supporting materials). These primary vacancies act as seeds for subsequent atom removal and finally for the creation of dislocation lines. Combining the higher knock-on cross sections for additional vacancies created adjacent to the first vacancy sites with the low concentration of primary vacancies, an overall emission of a few tens of atoms from the tube is estimated during each irradiation event. We note that thermal vacancy migration or spontaneous emission of atoms is unlikely at room temperature for C and BN systems.\cite{Krasheninnikov-05,Zobelli-07-1} This is the reason why the tube shaping is stable and why kinks are only obtained in the regions the electron irradiation conditions were optimized for.

\begin{figure}[t]
\centering
\includegraphics[width=\columnwidth]{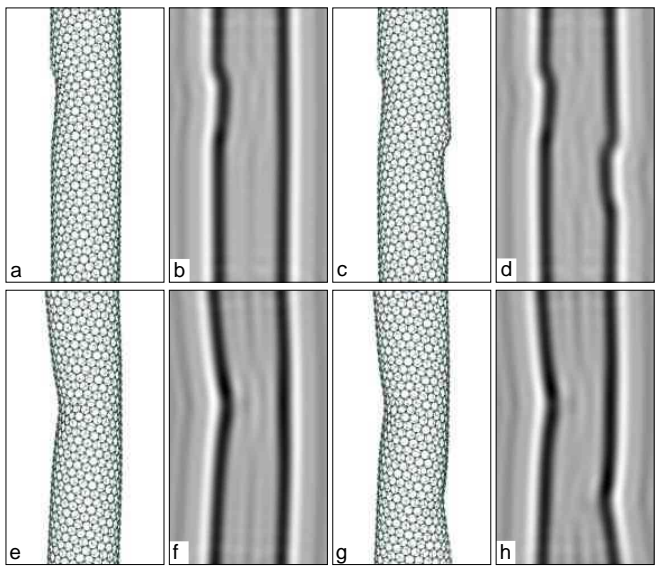}
\caption{Relaxed structure and respective STEM bright field simulated images for different dislocation lines in a (20,5) single walled carbon nanotube. Structures a and e correspond to a single dislocation line made of 12 missing atoms with different orientations. Structures c and g corresponds to the presence of two dislocations in the tube.}
\label{immsim}
\end{figure}

The introduction of a dislocation line produces a shortening of the tube whose length corresponds in a first approximation to the component of the Burger's vector along the tube axis. The bending of the tube visible in the microscopy images is then associated with shrinkage of one of the tube sides, and its magnitude depends on the orientation of the dislocation. 
This behaviour is illustrated by the example of figure \ref{immsim}. 

Considering a (20,5) chiral carbon nanotube, whose diameter is compatible with the tube of Fig. \ref{kinkseries}, the tube structures with differently oriented dislocation lines have been relaxed by DFTB and  microscope images have been simulated.
In figure \ref{immsim}.a the removal of twelve atoms along a direction of 10.95$^{\circ}$ with respect to the tube axis does not produce any visible bending of the tube (figures \ref{immsim}.b) but two separated kinks are clearly visible in the left wall.
When two dislocation lines are present in the tube, with a separation of several nanometers along the tube axis, the result is quite similar. No substantial bending of the tube is obtained and the dislocations are only visible though extended defects on the right and left part of the tube (figures \ref{immsim}.c,d).
Removing the same number of atoms along a different direction of 79.05 $^{\circ}$, closer to the normal of the tube axis, gives a larger shortening of one side of the tube, which corresponds after relaxation to a bending of the tubular structure (figures \ref{immsim}.e). The simulated microscope images with one or two dislocation lines (figures \ref{immsim}.d,f) reproduce well the behavior observed in the experimentally irradiated nanotubes (figures \ref{kinkseries}.b,c), demonstrating that few tens of atoms have been removed by the electron beam.

\begin{figure}[htb]
\centering
\includegraphics[width=0.8\columnwidth,bb= 0 0 255 590]{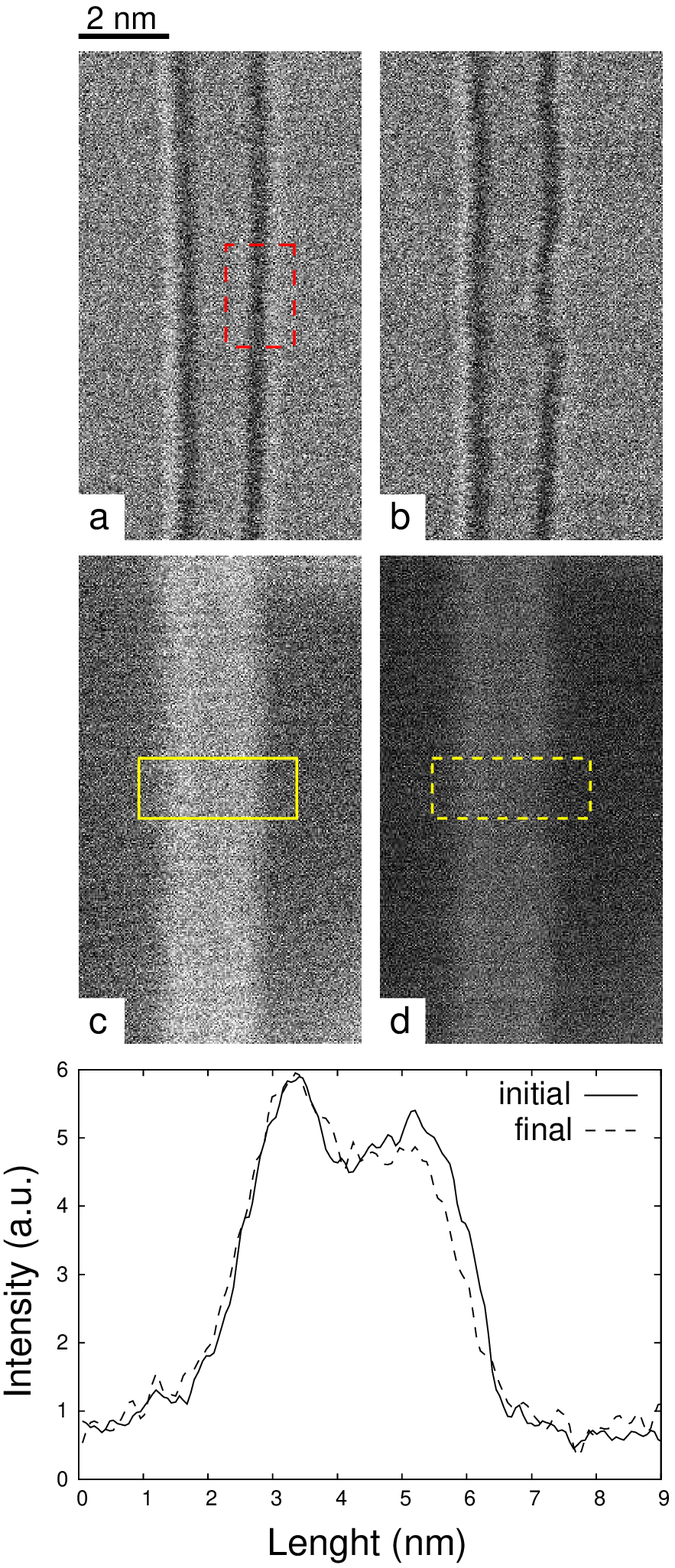}
\caption{Localised irradiation of a single walled BN nanotube. (a-b) Experimental STEM bright field images before and after the irradiation cycle of the nanotube. (c-d) Corresponding STEM dark field images before and after the irradiation cycle. Lower part of the figure, STEM dark field profile of the nanotube showing tube thinning on the right.
From the difference of the profile integrals it can be estimated that around 4\% of the atoms have been sputtered corresponding to the creation of around 40 vacancies.}
\label{BNseries}
\end{figure}

Similar irradiation conditions can be used in the shaping of single walled boron nitride nanotubes. Compared to carbon irradiation, defects in BN nanotubes appear at lower irradiation energies.\cite{Zobelli-07-2}  An irradiation beam energy of 80 KeV, slightly above the electron irradiation energy threshold, has thus been chosen in order to control the irradiation damage.  Figure \ref{BNseries}.a presents a single walled boron nitride nanotube before irradiation; figure \ref{BNseries}.b is taken after 60 s of irradiation, the dashed red rectangle in fig. \ref{BNseries}.a representing the $2 \times 3$ nm$^{2}$ irradiated area. After the irradiation cycle, the tube diameter appears locally reduced in the bright field image and two kinks appear in the tube walls. 

STEM annular dark field images give a direct correlation between the local intensity of the image and the local atomic density of the sample and can be used to quantify the loss of atoms (figures \ref{BNseries}.c and \ref{BNseries}.d). The intensity profiles obtained across the tube before and after the irradiation are legible in the lower part of the figure \ref{BNseries}. After the irradiation cycle is applied to the right side of the wall, the profile shows a decrease in intensity and the tube diameter shrinks from 2.0 nm to 1.8 nm. The profile integral beeing proportional to the number of atoms, an estimate of around 4\% of the atoms, corresponding to a few tens of atoms, have been sputtered from the tube during the irradiation cycle. This value is in agreement with the theoretically expected numbers of vacancies. 

Similarly to single walled carbon nanotubes, it is possible to repeat the irradiation procedure in different sections of a BN tube in order to re-shape the nanotube with a nanometric periodicity (see Sec. III in supporting materials). Further examples of STEM dark field images for both carbon and boron nitride nanotubes are also given in Sec. IV of the supporting materials.

The experiments presented in this work give the first indications for a "nano-electron-lithography" of single walled nanotubes, a new top down approach to locally control their nanostructures. In particular, we demonstrate that electron irradiation optimised on the basis of knock-on cross sections calculated in a density functional based framework can be used to remove a series of a few tens of atoms in a periodic area along the tube axis. 
 The removal of atoms along a dislocation line introduces pentagon-heptagon defect pair and changes locally the tube chirality from (\textit{m,n}) to ($m\pm 1,n\mp 1$). 
For particular pairs of the Hamada indices \textit{m,n} the dislocation line can then change locally the electronic character of the tubes from semiconductor to metallic and vice versa.\cite{Chico-96,Ruppalt-07}.  It is currently impossible to have a precise control in the chiral index change since electron microscopes have not yet attained the capability of knocking out a specifically targeted atom. Nevertheless, the knock-on cross sections presented here can also be used as a "guide book" for that purpose, a unique way to generate a wide variety of carbon-based quasi-1D conductive systems, from quantum wells to nanodiodes.

\appendix
\section{Methods}
\subsection{Scanning transmission electron microscope}

Experiments have been performed in a VG-HB501 STEM equipped with a tungsten cold field-emission gun. The vacuum in the vicinity of the sample was around $5\cdot 10^{-8}$ torr and has been obtained with an oil-free pumping system.  High mechanical sample stability was obtained by a top-entry sample holder system. 
The beam convergence half-angle was set to about 7.5 mrad and the used pole piece has a spherical aberration of around 3.1 mm. For voltage energy of 100 keV, it corresponds to the formation of an electron probe of around 0.8 nm in diameter at the sample surface. Electron probe sizes were slightly degraded when lower voltages such as 60 or 80 keV were used. Bright field (BF) images have been obtained with a collection semi-angle of 1.25 mrad and dark field (DF) images have been obtained with collection angle between 25 and 200 mrad. 
Irradiation conditions to obtain tube shaping are discussed in the body text. For BF/DF imaging the acquisition time was limited to 1-2 seconds and when needed beam blanking was used before the sample in order to limit the electron irradiation required to image the nanotube. In such conditions the statistical creations of vacancies during the imaging procedure of the tube was calculated to be around 0.1 over the full section of the imaged nanotube.  
Electron beam currents have been calibrated by a direct measurement of the current inside the drift tube of an EELS spectrometer using a Keithley pico-amperometer. 
Carbon SWNT nanotubes have been synthesized by the CVD process and are of commercial origin (Thomas Swan \& Co. Ltd). Boron nitride nanotube synthesis is described elsewhere\cite{Lee-01} and tubes have been provided to us by R. Arenal de la Concha and A. Loiseau (LEM-Onera, France).

\subsection{Calculations}

Knock-on cross sections have been calculated in a similar manner as described in detail in reference \cite{Zobelli-07-2}:
Atom emission occurs for scattering angles where the energy transferred to the nucleus is higher than a certain escaping energy. Firstly, maps of the emission energy threshold as a function of the emission angle have been obtained through extended density functional based  molecular dynamics simulations.\cite{Zobelli-07-2,Krasheninnikov-05}. 
Molecular dynamics has been performed using the density functional tight binding (DFTB) method simulating the local environnment of carbon nanotubes by a planar graphene sheet. 
Details of the DFTB parameters can also be found in the reference\cite{Zobelli-07-2}.
The total knock on cross sections have then been calculated by integration of the Mott cross section\cite{Mott-29,Mott-32,McKinley-48} for the set of angles satisfying the emission conditions.

In the microscopy image simulations section structural optimisations  have been conducted in the framework of the density functional tight binding theory\cite{Porezag-95,Seifert-96} using the DFTB+ code.\cite{Aradi-07} 
We have considered a (20,5) single walled carbon nanotube with a tube diameter of 1.8 nm.
In order to allow the bending of the tube calculations have been performed in a cluster mode using models containing up to 1800 atoms. Dangling bonds at open caps have been saturated by addition of hydrogen atoms. 

Microscopy simulated images have been obtained using the multi-slice simulation method included in the TEMSIM packages.\cite{Kirkland-98}
 
\section{Acknowledgments}

We would like to thank R. Arenal de la Concha and A. Loiseau for having kindly provided the BN nanotubes samples and M. Tenc\'e for the development in the STEM probe scanning system. A.Z. acknowledges the New Fullerene like materials network, contract HPRNCT-2002-00209, and the Enabling Science and Technology for European Electron Microscopy network, contract ESTEEM-0260019, for financial support.

\end{document}


\title{Supplementary informations for:\\\Large{Shaping single walled nanotubes}}

\author{A. Zobelli}
\email{zobelli@lps.u-psud.fr}
\affiliation{Laboratoire de Physique des Solides, Univ. Paris-Sud, CNRS, UMR 8502, F-91405 Orsay Cedex, France}

\affiliation{Technische Universit\"{a}t Dresden,
Instit\"{u}t f\"{u}r Physikalische Chemie und Elektrochemie, D-1062 Dresden,
Germany}

\author{A. Gloter}
\affiliation{Laboratoire de Physique des Solides, Univ. Paris-Sud, CNRS, UMR 8502, F-91405 Orsay Cedex, France}

\author{C. P. Ewels}
\affiliation{Universit{\'e} de Nantes, 
Nantes Atlantique Universit{\'e}s, CNRS,
Institut des Mat{\'e}riaux Jean
Rouxel (IMN), UMR6502, BP32229, F-44322 Nantes Cedex 3, France}

\author{C. Colliex}
\affiliation{Laboratoire de Physique des Solides, Univ. Paris-Sud, CNRS, UMR 8502, F-91405 Orsay Cedex, France}

\maketitle

\section{Total knock-on cross section: energy and angular dependence}

\begin{figure}[h]
  \hfill
  \begin{minipage}[t]{.45\textwidth}
    \begin{center}  
\includegraphics[width=0.7\linewidth,bb=0 0 484 720]{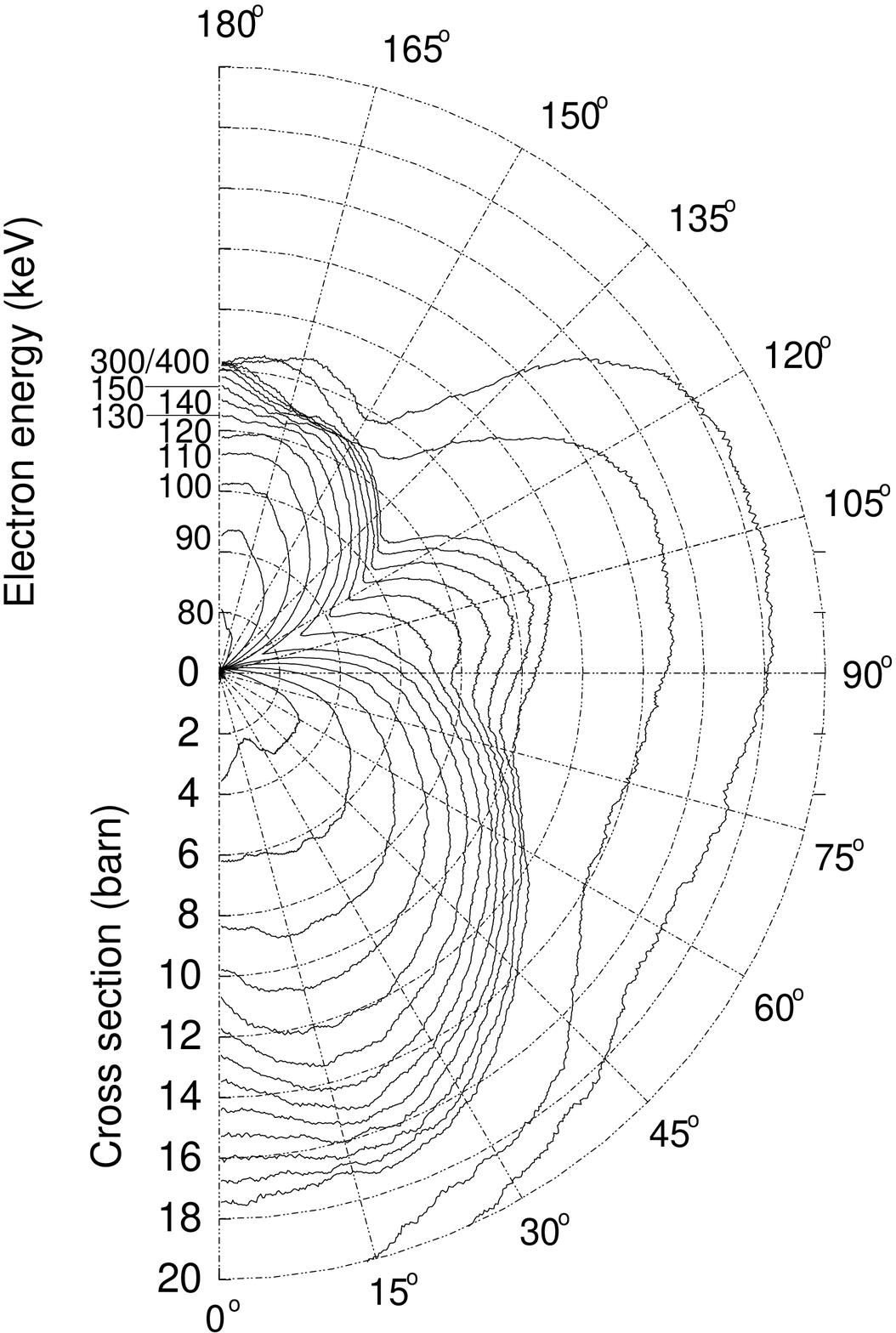}
\caption{Total knock-on cross section for a double coordinated atom neighboring a monovacancy as function of its position $\alpha$ around the tube circumference.}
\label{cross-monovac}
    \end{center}
  \end{minipage}
  \hfill
  \begin{minipage}[t]{.45\textwidth}
    \begin{center}  
\includegraphics[width=0.7\linewidth,bb=0 0 484 720]{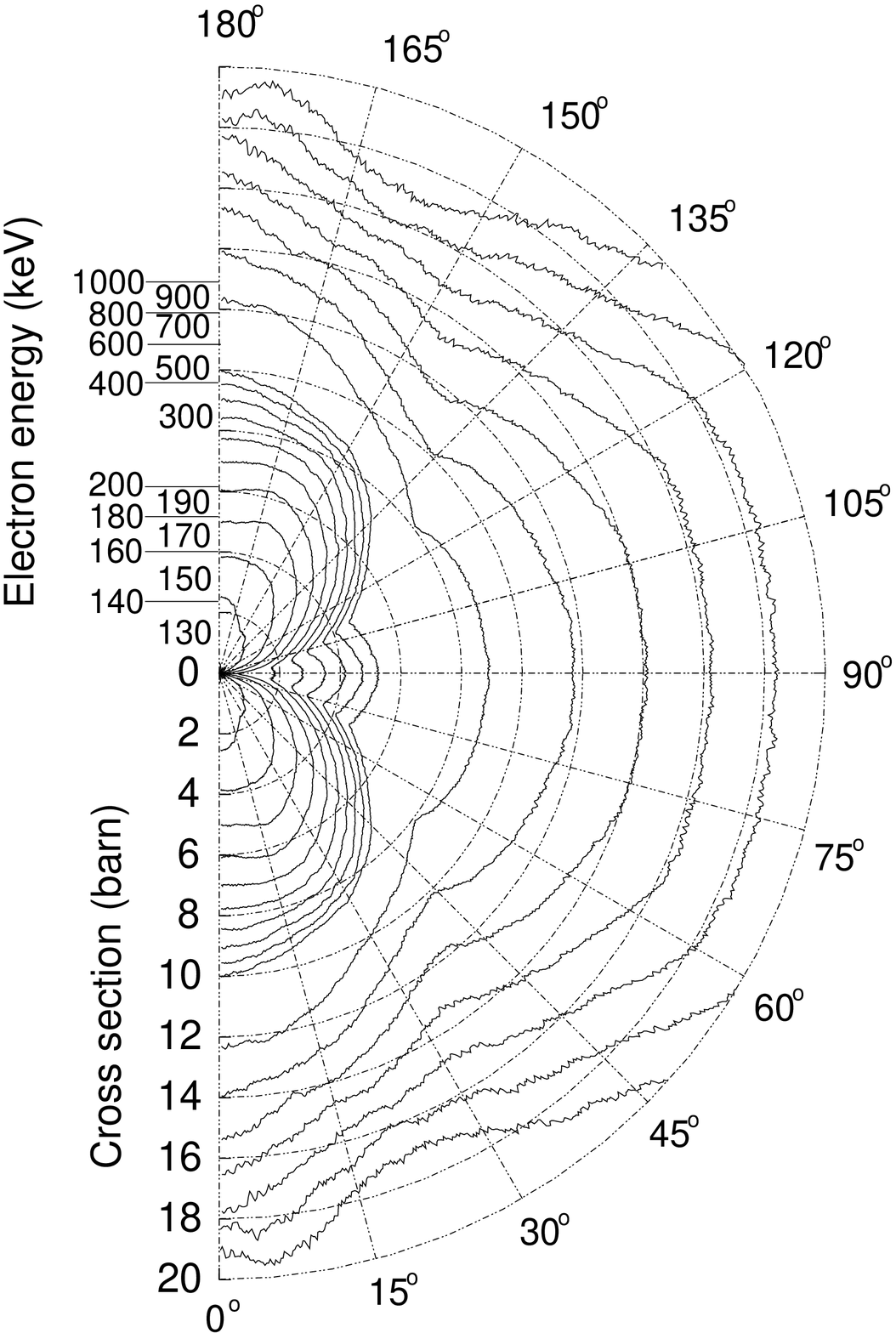}
\caption{Total knock-on cross section for a carbon atom neighboring a divacancy as a function of its position $\alpha$ around the tube circumference.}
\label{cross-divac}
    \end{center}
  \end{minipage}
  \hfill
\end{figure}

The map of the total knock-on cross section for atoms in a perfect carbon nanotube as function of the incident electron energy and the position of atoms around the tube circumference has been presented in Ref. \cite{Zobelli-07-2}.
Here we report analogous maps for the emission of carbon atoms in defective nanotubes: for a double coordinated atom neighboring a monovacancy in S. \ref{cross-monovac} and for a carbon atom neighboring a divacancy in S. \ref{cross-divac}.

The asymmetry between the upper and lower part of the tube in S. \ref{cross-monovac} is due to the outward movement of the doubly coordinated atom close to the vacancy, which makes emission into the tube cavity more difficult. The asymmetry of this defective structure modifies the angular dependence of the total knock-on cross section as a function of the orientation of the vacancy axis with respect to the tube axis. Nanotubes with different chiralities, on which the vacancy symmetry plane is oriented in different directions, would have slightly different maps of the total cross section.

\section{Total number of primary vacancies \label{primaryvac}}

\begin{figure}[bth]
\centering
\includegraphics[width=0.5\columnwidth]{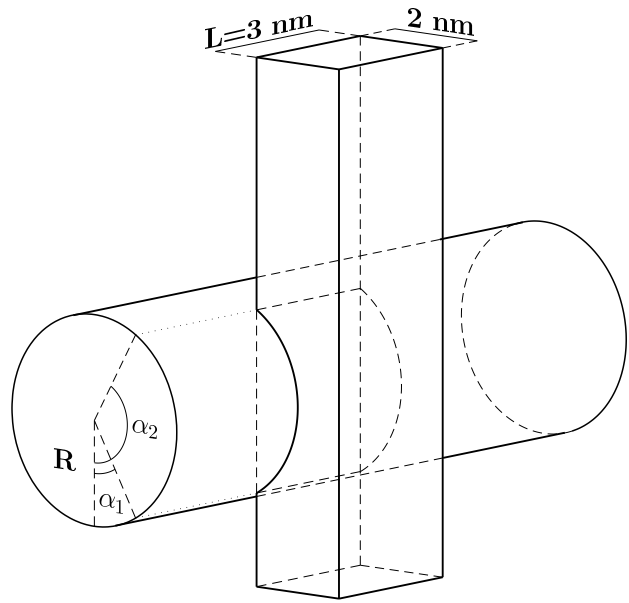}
\caption{Schematic representation of the irradiation geometry of a nanotube. The electron beam incidence direction is orthogonal to the tube axis.}
\label{tubeirrgeom}
\end{figure}

Transmission electron microscopy experiments are usually performed with nanotubes deposited onto a lacey carbon grid placed perpendicular to the TEM axis. Electron irradiation is then primarily performed in a non-tilted geometry where the tube axis lies perpendicular to the direction of the electron beam (see S. \ref{tubeirrgeom}). In this configuration the position of the atoms around the tube circumference can be identified using the angle $\alpha$ defined by the direction of incidence of the electron  and the local normal to the tube wall.
The number of events N which occur at a position $\alpha$ on the tube for an irradiation time $t$ can be obtained by:

\begin{equation}
N=jRLt\int_{\alpha_{1}}^{\alpha_{2}}\sigma(\alpha)\rho|\cos(\alpha)|\mathrm{d}\alpha
\label{integral}
\end{equation}

where $\rho$ is the atom density of a graphene plane, $j$ is current density, $R$ is tube radius and $L$ the illuminated length along the tube axis. $\rho$ represents then the atom surface density. The intersection between the electron beam irradiation zone and the nanotube is defined by the two integration limit polar angles $\alpha_1$ and $\alpha_2$.
In the experimental set-up the operating current density in the microscope was fixed at about $150 \cdot 10^{28}$ e$^{-}$/(s$\cdot$nm$^2$). We consider an atom surface density of  $9.68$ Atom/nm$^2$ for a graphene plane and an effective illumination area defined by an illuminated length L=3 nm  and the two polar angles $\alpha_1=60^{\circ}$  and  $\alpha_2=120^{\circ}$. Under these experimental conditions a total number of 2.74 vacancies are generated during an exposure time of $t=60$ s.

%
%
%
%

\section{BN single walled nanotube shaping}

\begin{figure}[bt]
\centering
\includegraphics[width=0.9\columnwidth]{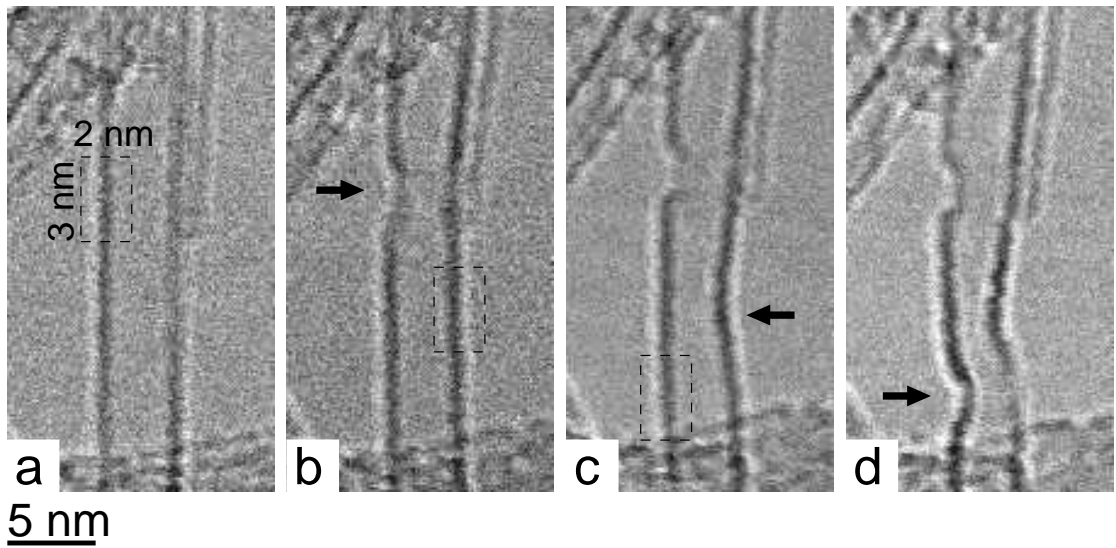}
\caption{Shape modification of a BN nanotube under localised irradiation.}
\label{BNseries}
\end{figure}

Irradiation experiments have been conducted on single walled BN nanotubes
 with an irradiation beam energy of 80 KeV, slightly above the electron irradiation energy threshold. 
Scanning regions are represented in S. \ref{BNseries} by the dashed rectangles and correspond to a $ 2 \times 3$ nm$^2$ area which partially illuminate the external section of the tube walls. Irradiation times (30 s) and electron beam current (140 pA) are identical to the values used for experiments in carbon nanotubes.
Similar to single walled carbon nanotubes, re-shaping of single walled BN nanotubes has been obtained by repeating irradiation cycles in different sections of the tube (Fig. S.\ref{BNseries}).

\begin{figure}[htbp]
\centering
\includegraphics[width=0.6\linewidth]{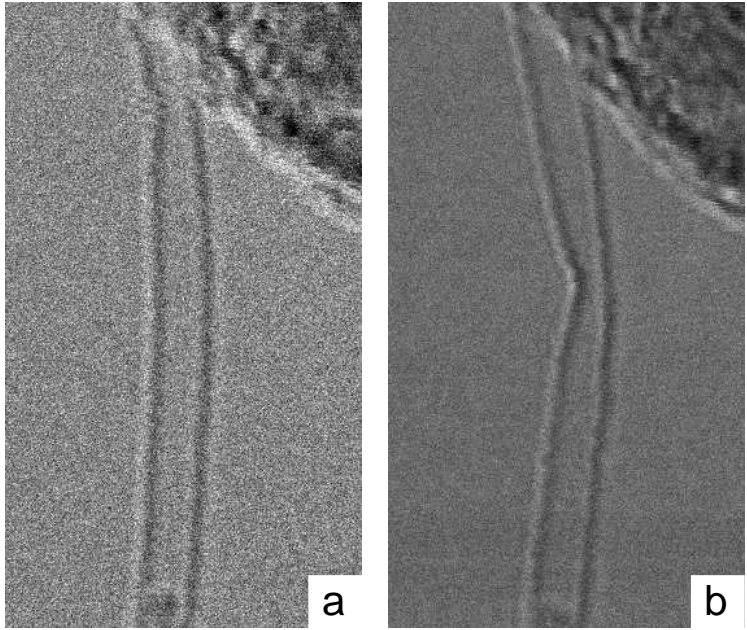}
\caption{BN nanotube bending obtained through localised electron irradiation}
\label{BNbend}
\end{figure}

Figure S.\ref{BNbend}  shows that, as for carbon nanotubes, local irradiation can produce a relevant bending of the tube. However this bending effect seems to be less common than for carbon nanotubes.  We propose that for BN nanotubes this effect is also related with the shortening of one of the tube wall sides.

\section{Annular dark field images}

\begin{figure}[p]
\centering
\includegraphics[width=0.6\linewidth]{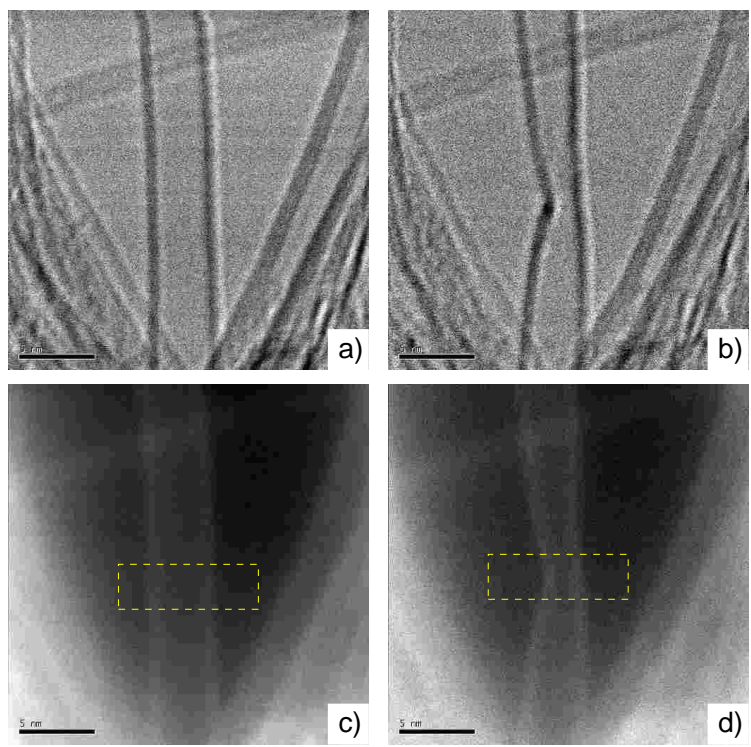}
\caption{Local irradiation of a large carbon nanotube: bright and dark field images. Dark field images shows a large tube diameter reduction associated with a tube bending.}
\label{serie8}
\end{figure}

\begin{figure}[p]
\centering
\includegraphics[width=\linewidth]{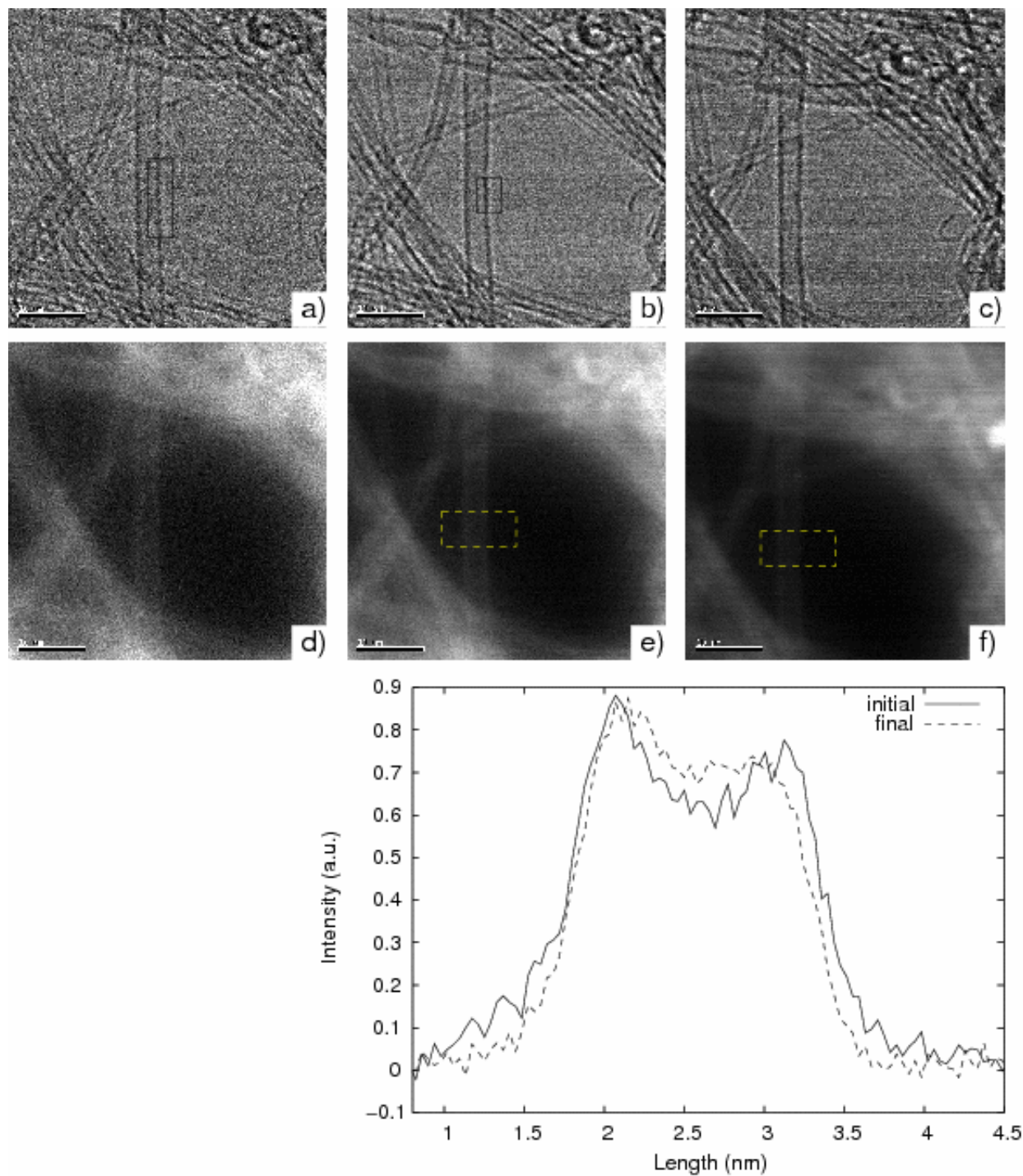}
\caption{Local irradiation of the right side of a single walled carbon nanotube: bright and dark field images. The lower part of the image shows the STEM dark field intensity profile in correspondence to the yellow rectangles in figures. The profile shows that atoms have been removed just from the irradiated side of the tube and we can estimate that 3\% of atoms have been sputtered from the selected area of the tube.}
\label{serie1}
\end{figure}

In this section  additional series of images on irradiation of carbon nanotubes obtained simultaneously in bright field and annular dark field (ADF) modes are shown. The series are presented at lower magnification compared to the images reported in the article  showing how the imaging process does not provoke relevant irradiation damage in the specimen. It clearly appears from the images (S. \ref{serie1}-\ref{serie8}) that non-irradiated zones are not modified within the image series.
This can be explained comparing the irradiation doses for imaging or irradiating the tubes.

Images are obtained by scanning a $20 \times 20$ nm$^2$ area for a period of 1-2 s (in S. \ref{serie1} the scanned area is $40 \times 40$ nm$^2$). Magnification is chosen in such a way that the tube occupies less then 1/10 of the image.
On the other hand, the irradiation cycles practiced to remove atoms are obtained by scanning $2\times 3$  nm$^2$ regions for a time of 30 s (where half of this region illuminates the tube). In this case the tube receives an electron dose which is 150 times the dose of the imaging process.

In Sec. \ref{primaryvac} we have calculated that, within the irradiation conditions usually adopted in this work, 3 mono-vacancies can be statistically created in the perfect tube side wall during the irradiation cycle, and that a few tens of atoms can then be further removed. In the imaging processes all the tube circumference is illuminated, thus the integration limits in Eq. \ref{integral} have to be $\alpha_{1}=0^{\circ}$ and $\alpha_{2}=180^{\circ}$. Repeating the integration with these new experimental conditions we find that around 0.125 primary vacancies are statistically generated during the imaging process in the whole perfect tube.

For thin specimens, an ADF signal is proportional to the projected mass thickness for example this can be used in to estimate the local density in organic thin films. In order to quantify the number of ejected atoms, we have measured the quantitative difference in the ADF profile  before and after atom ejection process, correcting the signal intensity with respect to the incident electron flux.  The initial numbers of atoms have then been estimated by computing the numbers of carbon or BN atoms in the tube (tube diameter can be accurately measured in ADF profiles). The number of ejected atoms is then obtained by using the mass loss coefficient obtained from the ADF. 

\begin{figure}[htbp]
\centering
\includegraphics[width=\linewidth]{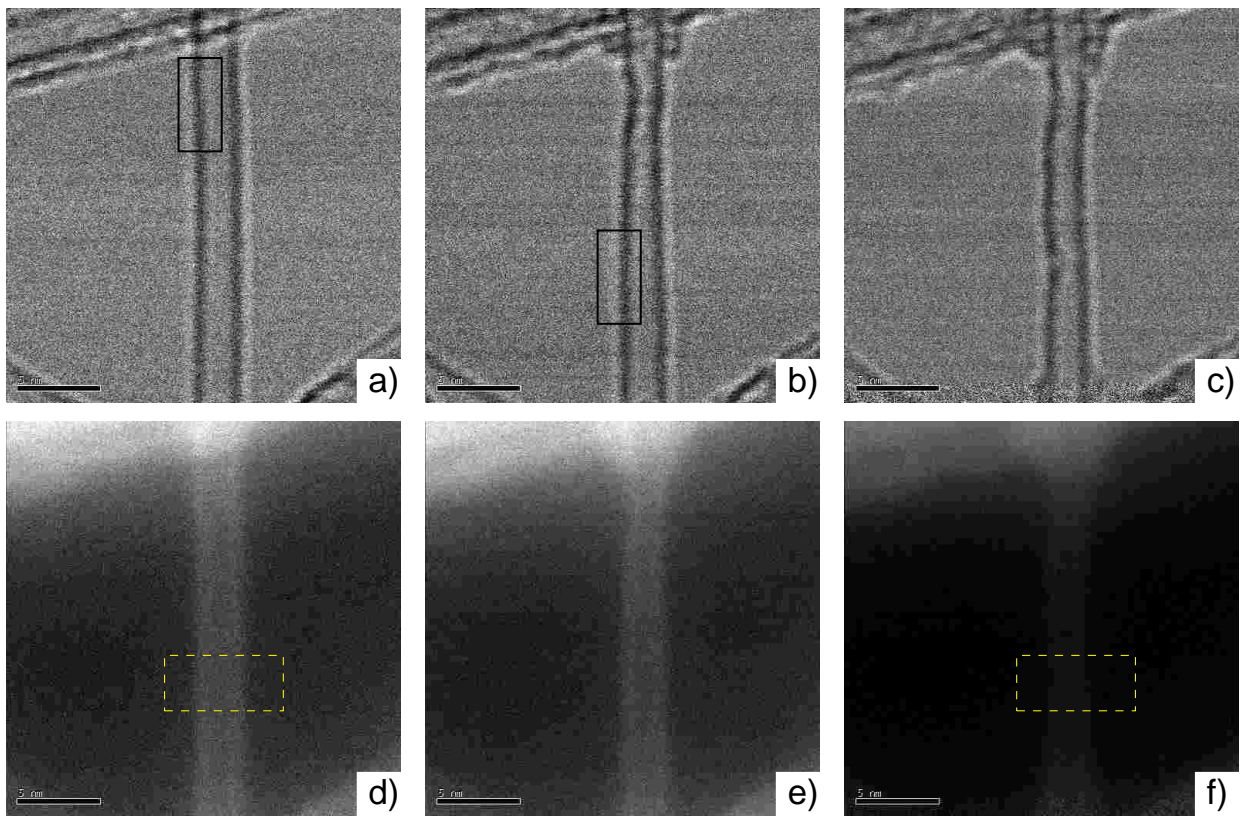}
\caption{Local irradiation of a single walled carbon nanotube: bright and dark field images. The intensity profile shows that atoms have been sputtered from the left side of the tube, with around 2\% of the atoms ejected from the selected area.}
\label{serie6}
\end{figure}

Figure S. \ref{serie8} shows for a carbon nanotube a large tube reduction associated with a slight tube bending. In figures S. \ref{serie1} and S. \ref{serie6} we report two series of irradiation of carbon nanotubes and respective ADF intensity profiles after successive irradiation cycles. We can estimate that around  3\% of carbon atoms  have been emitted in the selected area of figure S. \ref{serie1} and 2\% in the area of figure S. \ref{serie6}.

